\documentclass[aps,prl,showpacs,twocolumn,superscriptaddress]{revtex4}

\usepackage{epsfig}
\usepackage{graphicx}
\usepackage{amsmath}
\usepackage{amssymb}
\usepackage{amsfonts}
\usepackage{dcolumn}
\usepackage{color}

\newcommand{\sout}[1]{}

\begin{document}

\title{Coherent atomic soliton molecules for matter-wave switching}

\author{Chenyun Yin}
\affiliation{Department of Applied Mathematics and Theoretical Physics,
University of Cambridge, Cambridge, CB3 0WA, United Kingdom.}

\author{Natalia G. Berloff}
\affiliation{Department of Applied Mathematics and Theoretical Physics,
University of Cambridge, Cambridge, CB3 0WA, United Kingdom.}

\author{V\'{\i}ctor M. P\'erez-Garc\'{\i}a}
\affiliation{Departamento de
Matem\'aticas, E. T. S. I. I., Universidad de Castilla-La
Mancha 13071 Ciudad Real, Spain.}

\author{David Novoa}
\affiliation{\'Area de \'Optica, Facultade de Ciencias, Universidade de Vigo, As Lagoas s/n, Ourense, E-32004 Spain.}

\author{Alicia V. Carpentier}
 \affiliation{\'Area de \'Optica, Facultade de Ciencias, Universidade de Vigo, As Lagoas s/n, Ourense, E-32004 Spain.}

\author{Humberto Michinel}
 \affiliation{\'Area de \'Optica, Facultade de Ciencias, Universidade de Vigo, As Lagoas s/n, Ourense, E-32004 Spain.}

\begin{abstract}
We discuss the dynamics of interacting dark-bright two-dimensional vector solitons in multicomponent  immiscible bulk Bose-Einstein 
condensates. We describe matter-wave molecules without a scalar counterpart that can be seen as bound states of 
vector objects. We also analyze the possibility of using these structures as building blocks for the design of matter-wave 
switchers.  

\end{abstract}


\maketitle

\emph{Introduction.} Solitons are robust wave-packets able to maintain their shape when propagating in different media and 
under mutual collisions. The existence of such elastically interacting localized waves is an essential property of integrable nonlinear equations that have an infinite number of conservation laws \cite{1}. Integrable two-dimensional equations, such as Kadomtsev-Petviashvili, Davey-Stewardson, 
Zakharov-Manakov, 
sine-Gordon and others  have two-dimensional solitons with complex interactions. Some nonintegrable systems may have solitary waves -- localized coherent structures with almost elastic interactions. An interesting open question  is the construction of complex molecule-like coherent matter-wave structures, i.e. superpositions of solitons leading to stable bound states with 
 molecule-like behavior.
 
 In this paper we present nontrivial nonlinear phenomena in multicomponent Bose-Einstein condensates (BECs) described by coupled two-dimensional nonlinear Schr\"odinger equations (NLSEs). The balance of dispersion and nonlinear interactions in BECs leads to different types of nonlinear coherent excitations (see e.g. the experimental papers \cite{dark,bright,gap,vortices,shockwaves,vector,Faraday} or the reviews \cite{K1,K2}). 
 Soliton molecules have been considered in the propagation of optical beams in nonlinear media with saturable nonlinearities \cite{m1,m2,m3,m4}, it being very difficult to construct even metastable long-living soliton clusters with local interactions such as those present  in ordinary BECs. 

 Here we will construct soliton molecules 
using soliton-bubble bound states (i.e. two-dimensional extensions of the dark-bright soliton pair) 
as bricks to construct matter wave aggregates. These can also be viewed as a combination of the extension of a Kadomtsev--Petviashvili soliton \cite{manakov} and a bright NLS soliton or as a vortex pair superimposed with two density peaks. 

Matter-wave trains with a finite number of one-dimensional bright solitons
are stable due to the presence of the trap \cite{Gerdjikov}. 
However,  the idea does not work for higher dimensions due to the blow-up phenomenon \cite{bright}. 
With defocusing nonlinearities,  dark solitons always repel each other and 
cannot form bound states \cite{Gerd1}.

We will show how multicomponent homonuclear BECs in the immiscible regime 
allow for the construction of robust novel types of solitonic molecules. These matter-wave clusters 
display phase-dependent properties due to their coherent nature and can be used for constructing nonlinear matter-wave switchers.

\emph{Physical system and model equations.}  We will consider two-component BECs with atoms in two hyperfine states 
$|1\rangle$ and  $|2\rangle$ in the immiscible regime and consider droplets of atoms in 
component $|2\rangle$ to be phase separated from a component $|1\rangle$ assumed to have a much 
larger number of particles. When tightly confined 
 along one direction these systems are ruled in the mean field limit by
\begin{equation} \label{eq:6}
i \frac{\partial \psi_j}{\partial t} = -\frac{1}{2} \Delta \psi_j 
+ \left(\sum_{k=1,2}g_{jk}|\psi_k|^2-\mu+\delta_j\right) \psi_j,
\end{equation}
for $j=1,2$. Without loss of generality we work in
dimensionless units, with chemical potential  $\mu =1$, and 
$\delta_1=0$. Immiscibility implies that $g_{12}^2 > g_{11} g_{22}$. The normalization for $\psi_2$ is given by
 $\int_{\mathbb{R}} |\psi_2|^2 = 2 (a_{22}/a_0)N_2$,  where $a_{22}$ and $N_2$ are the s-wave 
scattering length and number of atoms in $|2\rangle$. Finally $a_0 =\sqrt{\hbar/m\omega_{\perp}}$ 
is the length-scale in which spatial units are measured.

\emph{Soliton molecules.}  In one-dimensional one-component NLS systems,  the repulsive nature of the interaction between dark solitons prevents them from generating bound states.
 The interactions between bright solitons in the absence of external effects (such as external confinement)  depend on the phase differences, $\Delta \phi=|\phi_1-\phi_2|$,  going from attractive for $\Delta \phi = 0$ to repulsive for $\Delta \phi = \pi$. 
A critical intermediate regime for $\Delta \phi$ exists in which unstable bound states can be constructed. 

A different possibility is constructing a vector object including a dark soliton in one component and a bright soliton in another 
component. This dark-bright pair could lead to a stable bound state with a second vector soliton of the same type when 
the repulsive  interactions between the dark components are balanced by the attractive interactions between the ``droplet-like" bright components \cite{KSh}.
However, when passing to higher dimensions the phenomenology changes 
essentially due to the fact that the transverse instability of the dark soliton 
leads to the formation of vortex pairs of opposite circulation. Therefore, the more natural building blocks for a bound state are vortices of the first component hosting a 
``droplet" of the second component. In the scalar case, all moving two-dimensional coherent states 
were found in Ref. \cite{jr}. These correspond to vortex pairs of opposite circulation and rarefaction pulses. A similar 
phenomenology arises in the two-component case in the miscible regime \cite{ngb}. As the velocity of the solitary wave increases, the
distance between vortices of opposite circulation decreases to zero. The solutions at even higher velocity are localized 
density perturbations without zeroes. In two dimensions the sequence of solutions terminates 
with solutions approaching zero energy and momentum as the velocity $U$ approaches the speed of sound. 
\begin{figure}
 \epsfig{figure=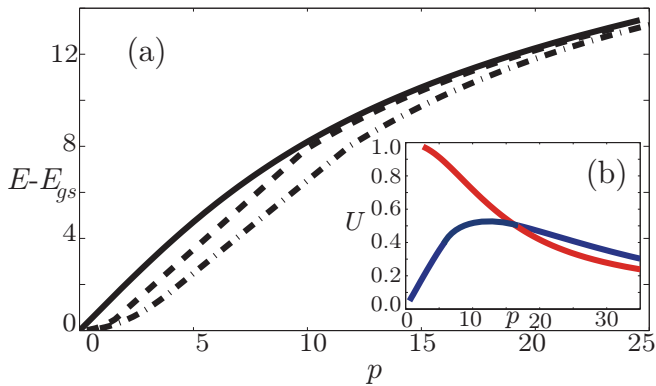, width=1\columnwidth}
\caption{(Color online)  (a) Energy ($E-E_{gs}$)-- impulse ($p=p_1+p_2$) dispersion curve of solitary
  wave solutions of Eqs. (\ref{ueq}) for $\alpha=1.2$ and various
 choices of $N_2$. The upper solid line corresponds to $N_2=0$, i.e. the 
 JR dispersion curve, shown for comparison. To ease comparison we subtract the ground state energy $E_{gs}$ in the plot. The dashed black line and the dashes-dotted line correspond to $N_2 = 8$ and $N_2= 20$, respectively.
 (b) Velocity as a function of momentum for $N_2= 20$; the red (monotonically decreasing) curve corresponds to the JR case; the blue (passing through the origin) curve correspond to nonzero $N_2$.}
\label{fig_disp}
\end{figure}

We will construct solitary waves with velocity $U$ along the $x-$direction 
in two-dimensional two-component BECs in the phase-separation regime as solutions of Eqs. \eqref{eq:6}
 in the frame moving with the disturbance:
\begin{subequations}
\label{ueq}
\begin{eqnarray}
i U \frac{\partial \psi_1}{\partial x} &=& \frac{1}{2}\nabla^2 \psi_1+
\left(1-|\psi_1|^2 - \alpha |\psi_2|^2\right)\psi_1,
\\
i U \frac{\partial \psi_2}{\partial x} &=&\frac{1}{2} \nabla^2 \psi_2+
\left(\Lambda-\alpha|\psi_1|^2 - |\psi_2|^2\right)\psi_2
\end{eqnarray}
\end{subequations}
together with the boundary conditions $| \psi_1|\rightarrow 1, \psi_2 \rightarrow 0,$ as $|{\bf x}|\rightarrow \infty.$ 
In the phase separation regime $\alpha=g_{12}/g_{11}=g_{12}/g_{22} > 1$. Here $\Lambda=\mu_2/\mu_1$ 
where $\mu_1$ and $\mu_2$ are the dimensional chemical potentials of $\psi_1$ and $\psi_2$.
We solve numerically the discretized version of Eqs. (\ref{ueq}) by a Newton-Raphson algorithm 
 combined with a secant algorithm to find
$\Lambda$ for a given constraint on  
$N_2=\int |\psi_2|^2\, dx\, dy$. We obtain a family of solutions characterized by the velocity
of propagation $U$, energy, $E$ and impulse ${\bf p}=(p_1+p_2,0)$, given by
\begin{subequations}
\begin{eqnarray}
E  & = &  \int \left[\frac{1}{2}|\nabla\psi_1|^2 +\frac{1}{2}|\nabla\psi_2|^2 +
\alpha|\psi_1|^2|\psi_2|^2 \right. \nonumber \\
& & +\left. \frac{1}{2}(1-|\psi_1|^2)^2 +
\frac{1}{2}|\psi_2|^4-\Lambda|\psi_2|^2 dxdy\right] ,\label{E} \\
 p_j  & = & \text{Im} \left[  \int_{\mathbb{R}^2}  \left(\psi_j^*-(2-j)\right)\frac{\partial \psi_j}{\partial x}   dxdy\right].
\end{eqnarray}
\end{subequations}
with $j=1,2$. The resulting families of
solutions are plotted in Fig. \ref{fig_disp} for various choices of $N_2$ 
together with the Jones-Roberts (JR) dispersion relations \cite{jr} for one-component condensates. For a given speed $U$, solitary solutions with higher $\alpha$  have lower energy and higher impulse. 

\begin{figure}
\epsfig{file=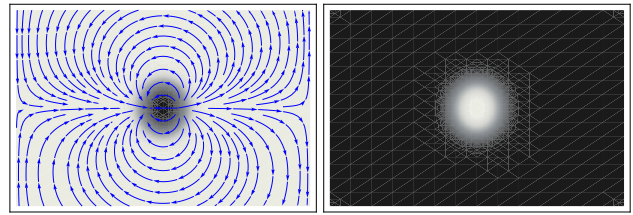,width=0.96\columnwidth}
\caption{(Color online) Density plots of the first (left) and second (right) components of a  stationary bound state solution for $U=0.1$, $N_2 = 80$. White color indicates maximum density. Streamlines of the first component are shown on the left panel. The spatial region shown is $60\times40$ healing lengths.}
\label{fig_density}
\end{figure}
In contrast with the JR solutions, there is a stationary solitary wave
with nonzero energy $E_{gs}$
corresponding to the ground state of the system with
all the mass of the second component forming a radially symmetric
``bubble'' in the center of the depleted first component.
As the
velocity increases from zero, the bubble becomes oblate in the direction of the motion with the velocity field of the 
first component being that of a dipole (see Fig. \ref{fig_density}). 

There is a point on the dispersion curve where the velocity
reaches its maximum -- the inflection point. As energy and momentum increase, the velocity decreases and the solutions become pairs of
vortices of opposite circulation in the first component with the second component filling the vortex cores. In general, bubble-like solutions for
small $E$ can be seen as a bound state of a JR rarefaction pulse and a mobile ``filling" of the second component. Fig. \ref{fig_disp}(b) shows 
that there is a maximum velocity for the propagation of these solutions (different from the sound speed). This is a signature of the mass of the 
second component, the heavier being the second component the smaller being this velocity.

\begin{figure}
\epsfig{file=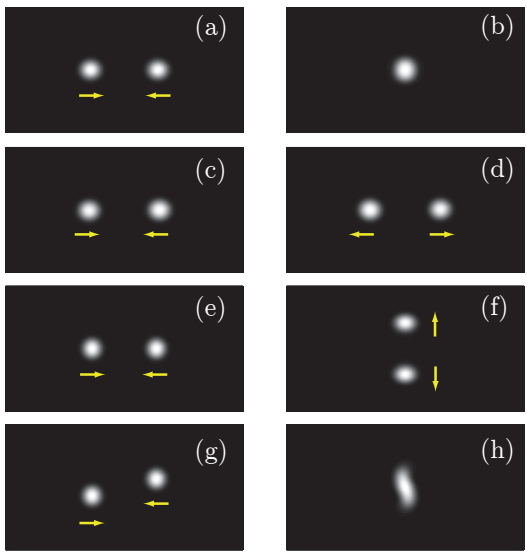,width= 0.8\columnwidth}
\caption{Density snapshots of the second component in coherent bubble-droplet pairs collisions, before (left column) 
and after (right column) the collision. (a-b) Identical incident bubbles with $U=0.1, N_2=40, p=6.36,  E=2.99$ form a bound state.  
(c-d) Incident bubbles with the same speed $U_1=U_2=0.2$ but different sizes emulate an elastic collision (left bubble: 
$N_2=40, E=5.6, p=14.9$, right bubble: $N_2=50, E=7.7, p=20.1 $).  (e-f) Identical bubbles (parameters being 
$U=0.2, N_2=40, E=5.6, p=14.9$) emulate the collision of two pairs of well-separated vortices.
(g-h) Identical bubbles (same as those in third row) collide with an offset of 8 dimensionless units and form a bound state. The spatial region spanned is $x \in [-50,50], y \in [-30,30]$.}
\label{fig_collision}
\end{figure}

We have simulated the evolution of two bubbles set on a colliding course. Initially they are separated by a large distance, so
that individually they are accurately represented by the solutions we found. Several possible outcomes of such collisions are 
summarized in Fig. \ref{fig_collision}. Almost identical slow colliding bubbles may form a bound stationary state  even when 
they collide with an offset. Bubbles moving with large velocities may scatter at $\pi/2$ angle resembling the collision of two pairs 
of well-separated vortices of opposite circulation. Almost elastic collisions between these structures were observed  when the 
velocities or masses of the bubbles were very different. A bound state is more likely to  be formed when bubbles have similar 
phases of the second component and move slowly. In such collisions, a small fraction of the mass is emitted as sound waves. 
The outgoing bubbles are solitary waves as verified by energy-impulse calculations.

Our simulations of Eqs. (\ref{eq:6}) were done using two high order schemes: a fourth-order in space finite differences with a 
fourth-order Runge-Kutta scheme in time (FD) and a Fourier pseudospectral method (FS) \cite{PG}, with  $\Delta x= \Delta 
y =0.25$ and $\Delta t=0.01$ (FD), $0.005$ (FS).

\emph{Dynamic molecules.} These solitonic molecules may have complex oscillatory internal modes: the states formed by vortices 
with a bright filling, exchange parts of the mass leading to a periodic beating, as we will show simulating the evolution of initial conditions:

\begin{subequations}
\label{input}
\begin{eqnarray}
\psi_1&=&\text{tanh}\left( \frac{r_+}{w_d}\right)e^{-i\theta_+}\cdot\text{tanh}\left( \frac{r_-}{w_d}\right)e^{i\theta_-},
\\
\psi_2&=&e^{-\left( \frac{r_+}{w_b}\right)^2}.
\end{eqnarray}
\end{subequations}

\begin{figure}
\epsfig{file=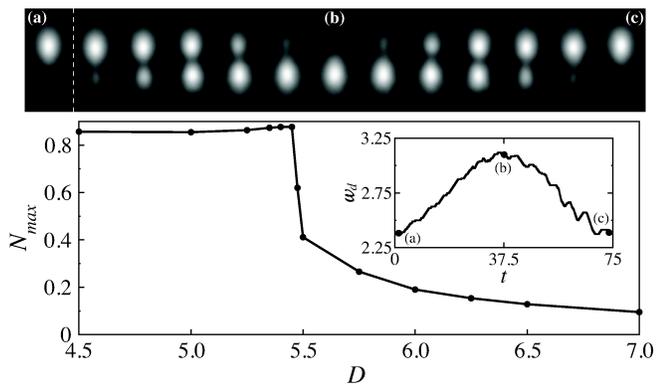,width=1\columnwidth}
\caption{Top: Propagation of a vortex pair of opposite circulation 
in component $|1\rangle$ with a bright soliton in component $|2\rangle$ nested into one of the cores. The pseudocolor 
plot shows the density of the bright component.
The snapshots are taken in the center-of-mass system with a separation of  6.25 time units, each showing a spatial window of 
$23.5\times9.5$ units of a larger simulation region (100$\times$100). 
The initial separation between both cores is $D=4.5$. Bottom: Switching curve $N_{max}$ vs. $D$. Inset: Evolution 
of $w_d$ for the lower vortex core in the simulation displayed in the top.}
\label{beating_long}\end{figure}

Being $r_\pm=\sqrt{(x\pm \frac{D}{2})^2+y^2}$, $\theta_\pm=atan\left(y/\left(x\pm \frac{D}{2}\right)\right)$, $w_d$ and $w_b$, respectively, the widths of the vortex cores and 
the "bright" component of these multidimensional dark-bright solitons and
$D$ the separation between the centers of the "dark" parts. The choice of Gaussian and hyperbolic tangent profiles for the bright component and 
vortices, respectively, is more realistic from the experimental point of view than numerically calculated eigenstates. 
We consider $N_2=5$ as number of particles of the bright component. We also impose $w_b=1.0$ 
and that the diameter of the bright part fits exactly within the vortex size at one half of the maximum amplitude, which 
yields $w_d=2.3$, calculated by the condition: $|\psi_1\left(\frac{w_d+D}{2}\right)|^2=|\frac{1}{2}\psi_1(\infty)|^2$. 
We have chosen $g_{ij}=-1$, with $j=1,2$, to illustrate that the results are valid even in the 
boundary of the immiscibility region. In this configuration the two 
nonlinear structures move along a rectilinear path, keeping the distance $D=4.5$ fixed. The bright component of the pair first tunnels to the empty vortex core and then beats periodically between the two cores similarly to 
the bright soliton oscillation in a double-well potential (see Fig. \ref{beating_long} (top)). The vortex cores significantly change their size along the 
whole dynamics, the effective potential wells being dynamically modified along the direction of
propagation owing to the inter/intra-component interactions (see the inset in Fig. \ref{beating_long}). Initially (point (a)), the vortex is empty, and therefore 
features a minimum width. As the bright component starts to fill up 
the vortex core, it becomes broader owing to the repulsive interaction between different atomic species, reaching 
a maximum width for $t=37.5$ (point (b)). When the bright component 
 in a vortex decreases, it narrows again (see point (c) for $t=75$).

The bottom plot in Fig. \ref{beating_long} shows the variation of the maximum of the  bright population 
within the initially-empty core $N_{max}$  (normalized to the total number of particles in the bright component) 
as a function of $D$. As it can be appreciated in the graph, when the cores are close enough, most
particles oscillate between both cores. A small fraction of particles remains in the empty 
vortex, because of the overlapping between modes in both cores. As $D$ increases over 
a certain threshold the tunneling between cores is suppressed due to the 
nonlinearity-induced asymmetry of the effective potentials and the  value of $N_{max}$ quickly decreases to zero. 
Thus, this {\it switching curve}, having an intermediate regime of partial tunneling in which is possible to 
control $N_{max}$ efficiently, allows for the design of a ``matter-wave switcher''. It is also possible to control
both $N_{max}$ and the beating period by slightly tuning the interespecies 
coupling coefficients $(g_{12},g_{21})$, making the switcher completely reconfigurable.
Control of $g_{12}$ and $g_{21}$ has been demonstrated
previously in experiments \cite{thal}.
The beating persists for times longer than the condensate lifetime, indicating 
that the {\em nonlinear matter-wave switching} process is very robust.

\begin{figure}
 \epsfig{file=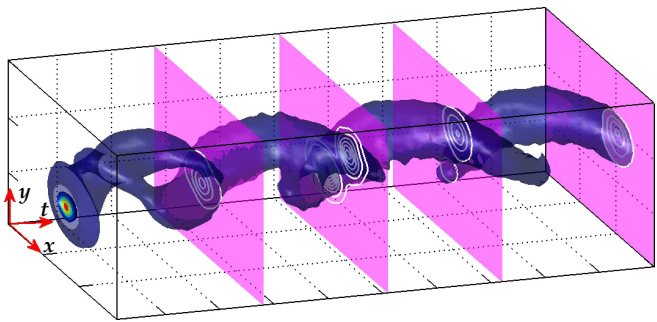,width=1\columnwidth}
\caption{(Color online) Density isosurface plot of the beating process of the bright field 
hosted by two equally-charged vortices. Dark cores rotate keeping the distance to its 
center of mass unaltered. A periodic beating of the 
bright component (blue surface) is observed in between both cores, as it can be inferred from the white contours within the inner 
slices. In this plot, $D=3.5$, the grid size is $100\times100$, the spatial range is 
$x,y\in[-10,10]$, $\alpha =1$ and the time interval $t\in[0,150]$.} 
\label{beating_rot}
\end{figure}

A second interesting configuration corresponds to a pair of vortices of {\em equal} charges and the same initial conditions described previously in the discussion of the results shown in Fig. \ref{beating_long}. For this system, the vortices orbit following circular trajectories while the bright component tunnels between the vortex cores and thus follows a spiralling trajectory 
(see Fig. \ref{beating_rot}). This beating is very stable and does not affect the circular vortex trajectories.

\emph{Conclusions.} We have discussed the dynamics of interacting bubble-droplet pairs in
quasi-two dimensional immiscible BECs. We have found novel types of robust 
bubble-like solitons without a scalar counterpart that can be used to construct coherent atomic soliton molecules.
The study of the dynamics of these stable objects has revealed the possibility of constructing \emph{nonlinear matter-wave switchers}. Our ideas could be tested in future experiments.

\acknowledgments

\emph{Acknowledgement.} This work has been supported by grants
FIS2007-29090-E (Ministerio de Ciencia e Innovaci\'on, Spain), PGIDIT04TIC383001PR (Xunta de Galicia), and PEII11-0178-4092 (Junta de Comunidades de Castilla-La Mancha)
  NGB acknowledges EU FP7 ITN project CLERMONT4.

\end{document}